\documentclass[sigconf]{acmart}

\usepackage{booktabs} 

\setcopyright{rightsretained}

\acmYear{2018}

\begin{document}
\title{Testing Alignment of Node Attributes with Network Structure Through Label Propagation}

\author{Natalie Stanley}
\affiliation{
\institution{The University of North Carolina at Chapel Hill}
}
\email{NatalieStanley1318@gmail.com}

\author{Marc Niethammer}
\affiliation{
\institution{The University of North Carolina at Chapel Hill}
}
\email{mn@cs.unc.edu}

\author{Peter J. Mucha}
\affiliation{
\institution{The University of North Carolina at Chapel Hill}
}
\email{mucha@unc.edu}

\begin{abstract}
Attributed network data is becoming increasingly common across fields, as we are often equipped with information about nodes in addition to their pairwise connectivity patterns. This extra information can manifest as a classification, or as a multidimensional vector of features. Recently developed methods that seek to extend community detection approaches to attributed networks have explored how to most effectively combine connectivity and attribute information to identify quality communities. These methods often rely on some assumption of the dependency relationships between attributes and connectivity. In this work, we seek to develop a statistical test to assess whether node attributes align with network connectivity. The objective is to quantitatively evaluate whether nodes with similar connectivity patterns also have similar attributes. To address this problem, we use a node sampling and label propagation approach. We apply our method to several synthetic examples that explore how network structure and attribute characteristics affect the empirical $p$-value computed by our method. Finally, we apply the test to a network generated from a single cell mass cytometry (CyTOF) dataset and show that our test can identify markers associated with distinct sub populations of single cells. 
\end{abstract}


\maketitle

\section{Introduction}
Community detection in networks is a common pursuit that seeks to partition the network's nodes into sets of structurally coherent groups, where members of a \emph{community} have strong similarity in connectivity patterns \cite{muchacommunity,fortu2,shaicase}. While the identification of communities based solely on the network's adjacency matrix is straightforward, the implications of having node attributes and how to integrate that into the community detection problem is not as well understood. We refer to a \emph{structural community} as a community identified according to only the adjacency matrix, while we define an \emph{attribute community} as a community that has been identified using the attribute information. Recently, there have been numerous approaches extending common community detection techniques to attributed networks \cite{hric,peel2017ground,ilouvain,cesna,clauset,perozziAttribute}. While each of these methods provide extensions to a variety of community detection approaches, they also differ in their assumption about the dependence relationships between the attributes and connectivity. On one hand, it seems reasonable to assume that members of a structural community should be highly similar in attribute space. However, work by Newman \emph{et al.} \cite{clauset} and Peel \emph{et al.} \cite{peel2017ground} have provided phenomenal examples and insight into when this assumption could be invalid. 

In this work, we seek to develop a test that returns a statistic reflecting how closely node attributes correlate with connectivity patterns. Our test is based on label propagation and ultimately returns an empirical $p$-value that can be interpreted as the significance of the relationship between network connectivity and node attributes. We demonstrate that the computed empirical $p$-value is meaningful with several synthetic examples and on a network representation of a single-cell mass cytometry CyTOF dataset. 

This paper is organized as follows: First, we describe the latest advances in attributed community detection. Next, we define our method and show several synthetic examples to evaluate the performance and meaningfulness of our computed empirical $p$-value. Finally, we apply our test to a single cell mass cytometry CyTOF dataset, identifying identify particular attributes that can distinguish populations of cells. 

\subsection{Community detection in attributed networks}
There are a variety of approaches for identifying structural communities based only on connectivity patterns such as probabilistic methods \cite{sbmOrig,bigclam}, quality function maximization \cite{newmanmodularity}, and higher-order motif-based clustering \cite{benson}. Most of these classes of methods have inspired extensions in attributed networks. Here, we discuss extensions to probabilistic and quality function maximization approaches. Note that throughout this text we use the phrases \emph{attributes} and \emph{metadata} interchangeably to be consistent with the terminology in the discussed references. 

\subsubsection{Probabilistic approaches}
Probabilistic community detection methods edge existence in a network based on latent community structure. After learning model parameters through likelihood optimization, samples generated from the model should align closely with the true underlying network. Two common approaches are the stochastic block model (SBM) \cite{sbmOrig} and the affiliation model \cite{bigclam}. 

The assumption of the stochastic block model is that nodes within a community are connected to nodes within and between communities in a characteristic way. Moreover, the objective in the model fitting and parameter inference of a stochastic block model in a network with $K$ communities is to learn the node-to-community assignments and the within and between community connection probabilities that maximize the model likelihood. The stochastic block model has been extensively studied in the literature and has been extended to attributed networks in at least four ways \cite{hric,peel2017ground,clauset,stanley2018}. Newman \emph{et al.} provided an extension to the stochastic block model capable of incorporating discrete or continuous metadata (attribute) information \cite{clauset}. Peel \emph{et al.} proposed the neoSBM \cite{peel2017ground}, which explores the effects of `fixing' and freeing nodes from their metadata label on the SBM inference. Along with this work, the authors developed BESTest, a statistical test to measure the relationship between a node's metadata label and community structure. Hric \emph{et al.} constructed a joint stochastic block model for both the attributes and metadata through a nonparametric, Bayesian framework \cite{hric}. They assessed the alignment of the attributes with the connectivity based on performance in link prediction tasks. Stanley \emph{et al.} introduced a version of the stochastic block model for networks with nodes having multiple continuous attributes \cite{stanley2018}.

Another useful probabilistic model for community structure is the affiliation model. This model assumes that nodes can be affiliated to multiple communities to varying extents \cite{bigclam}. Moreover, the edge between a pair of nodes is based on their similarity in community affiliations. A useful method for integrating multidimensional vectors of binary attributes was introduced by Yang \emph{et al.} in a method called CESNA \cite{cesna}, which modifies the affiliation model likelihood to incorporate this information. This is achieved by allowing the attributes and connectivity information to be modeled as conditionally independent, giving the node-to-community affiliations and feature importance weights for the attributes.

\subsubsection{Quality function maximization}
Quality function maximization methods have also been extended to attributed networks. When community detection is formulated with a quality function, the objective is to specify a null model for a network with no community structure and find the partition of nodes to communities that maximizes the difference from this null model. A standard quality function for communities is known as modularity \cite{newmanmodularity}. The state-of-the-art optimization heuristic for maximizing modularity is the Louvain algorithm \cite{blondel}. Work by Combe \emph{et al.} adapted the modularity to take into account multidimensional attribute vectors and optimized this quantity in a Louvain-style manner with I-Louvain \cite{ilouvain}. 

Recent work by Perozzi \emph{et al.} defines a modularity-inspired quantity known as \emph{community normality} \cite{perozziAttribute}. This measure prioritizes partitions where members of a community are very similar to each other in attribute space (and obviously in connectivity patterns). Further, members of a community are also expected to be different from nodes on the community boundary or in a different community. 

\subsection{Novelty and paper objectives}
The methods described above indicate great progress in the integration of attributes in community detection. However, there has not been much work focusing on the interplay between attributes and connectivity information and the extent to which these data should be combined. In this paper, we seek to develop a statistical test to evaluate the relationship between connectivity and attribute information. Our approach is meant to be generalizable to all networks and agnostic to the particular community detection algorithm algorithm applied to the network. Our method can also accommodate multiple attributes that can be either discrete, continuous, or a combination of both. After defining the method for implementing this test, we seek to systematically study the properties of our test and its empirical $p$-value output across various types of networks and attributes. Finally, we apply our tool to a single cell mass cytometry (CyTOF) dataset, where our results suggest that our method can successfully identify attributes with discriminative ability for distinguishing between communities.

\section{Methods}
Our method is built on label propagation (LP), where given a partially labeled network of $N$ nodes (i.e. only a subset of nodes have community assignments), the objective is to use this information to predict the community assignments of the unlabeled nodes \cite{LabProp}. In this work, we first \emph{label} the nodes according to their attribute information. We then take several sub samples of $l$ labeled nodes and use the prediction of the remaining $N-l$ unlabeled nodes as a proxy for how closely the attributes align with the network connectivity. In particular, we use a label propagation approach that returns a probability distribution for each node over each of the attribute-defined node-to-community assignments. We then quantify the uncertainty of the prediction with a simple entropy measure. In doing this, we assume that if the attributes are aligned with the network connectivity patterns, the entropy should be low. Alternatively, if attributes and connectivity are disparate, then predicting the unlabeled nodes will be difficult and entropy should be higher. 

\begin{figure}[h!]
\centering
\begin{center}
\includegraphics[width=.4\textwidth]{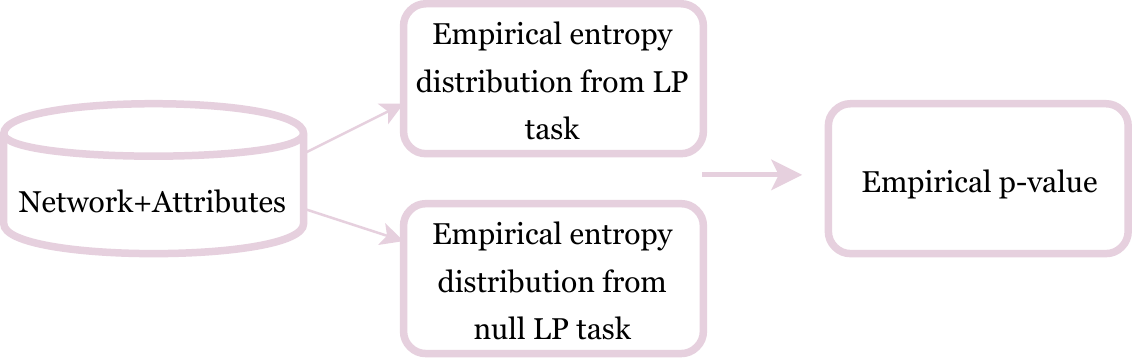}
\caption{Our test first labels the nodes according to the attribute-based node-to-community assignment, $\tilde{{\bf z}}$. Next, in each of of $T^{*}$ trials, a sample of $l$ labeled nodes is used as input to the label propagation task to predict the probability distribution over possible communities for the unlabeled $N-l$ nodes. The entropy of the node-to-community assignment probabilities is used to estimate how well the attributes align with connectivity. In each trial, $\tilde{{\bf z}}$ is permuted and subjected to the label propagation task to compute a `null' entropy value. This process is repeated $T^{*}$ times and the empirical $p$-value is calculated based on the overlap between the null entropy distribution and the empirical entropy distribution.}
\label{Overview}
\end{center}
\end{figure}

\indent As an overview of this process, we first label the nodes according to their attribute information. This can be achieved by classifying the nodes according to a single discrete value, or through simple clustering of the nodes based on their attributes. After obtaining a labeling of the nodes, we begin our label propagation and null label propagation tasks. For a large number of $T^{*}$ trials, we take a sample of $l$ nodes and their attribute-based labels and denote these nodes as \emph{labeled}. We then try to predict the labels of the remaining $N-l$ nodes, comprising the \emph{unlabeled} set, by propagating the labels outward. Since the label propagation method returns a probability distribution over possible community assignments, we can compute an entropy measure. Along with this true label propagation task, in each trial we also permute the labels of the nodes in our sample set to generate a null distribution of entropy values for the unlabeled nodes. Finally, the overlap between the null and empirical entropy distributions are used to compute a $p$-value. This process is outlined in Figure \ref{Overview}. We will now provided a detailed description of each step in this process. 

\subsection{Notation}
We first define some notation that assists in setting up this problem. For a network with $N$ nodes, we let ${\bf z}$ be the $N$-length vector of node-to-community assignments, based on only the network connectivity information given in the adjacency matrix, ${\bf A}=\{A_{ij}\}$. This implies that the $i$-th entry, $z_{i}$, gives the community assignment for node $i$. Alternatively, when nodes are labeled according to the attribute information, we denote their community assignments with $\tilde{{\bf z}}$. Finally, our permutation test involves taking a subset of nodes and their labels in $\tilde{{\bf z}}$ as the \emph{labeled} set to propagate labels outward to a set of \emph{unlabeled} nodes. We denote this distinction between the attribute-based partition of the labeled and unlabeled subsets of nodes by $\tilde{{\bf z}}^{L}$ and $\tilde{{\bf z}}^{U}$, respectively. Finally, we assume that each node has $p$ associated attributes, which are stored in the $N \times p$ matrix, ${\bf X}$. That is, the $i$th row of ${\bf X}$, $X_{i}$, gives the values of the $p$ attributes for node $i$. 

\subsection{Classifying Nodes}
The first step is to classify nodes according to attributes, denoted by $\tilde{{\bf z}}$. We assume some prior knowledge for the $K$, specifying how many communities are in the data. Hence, each $\tilde{z}_{i}$ takes on 1 of $K$ values. In the case where nodes are classified discretely, according to a single source of information, this labeling occurs without any effort. In the case where each node has multiple attributes, we have found that a simple clustering method, such as $k$-means works well. Because this first step of labeled the nodes is achieved through a clustering procedure, our approach can accommodate discrete and continuous attributes. 

\subsection{Sampling Nodes and Creating Entropy Distributions}
In the sampling step, for a large number of $T^{*}$ trials, we randomly select $l$ nodes, $\{L\}$, and their corresponding labels, $\tilde{{\bf z}}^{L}$. From here, we seek to use $\tilde{{\bf z}}^{L}$ and the network structure to predict the labels for the the remaining $N-l$ nodes in the unlabeled set, $\{U\}$. 

\indent After splitting all $N$ nodes into their labeled and unlabeled sets, we use the label propagation approach described by Zhu \emph{et al.} \cite{learning} to generate a probability distribution over the communities from $\tilde{{\bf z}}^{L}$ for each of the nodes in $\{U\}$. Ultimately under this LP approach, we seek to define the $N \times K$ matrix, ${\bf Y}$, where $Y_{ic}$ is the probability that node $i$ belongs to class $c$. We can split this matrix into two matrices, ${\bf Y}^{L}$ and ${\bf Y}^{U}$, containing the subset of rows corresponding to nodes in $\{L\}$ and $\{U\}$, respectively. Therefore, the label propagation task is to effectively estimate ${\bf Y}^{U}$. 

 To compute ${\bf Y}^{U}$ following the approach by Zhu \emph{et al.} \cite{learning}, we first use the adjacency matrix, ${\bf A}$ , to define and $N \times N$ transition probability matrix, ${\bf T}=\{T_{ij}\}$. Here, entry $T_{ij}$ gives the probability of jumping from node $j$ to node $i$. This is given by,
 
 \begin{equation}
 T_{ij}=\frac{A_{ij}}{\sum_{f=1}^{N}A_{fj}}.
 \end{equation} 
 
We then compute ${\bar{\bf T}}$, which is simply the row normalized version of ${\bf T}$. From here, ${\bar{\bf T}}$ is rearranged so that the first $L$ rows and columns correspond to the labeled nodes, and the remaining $N-L$ columns correspond to the unlabeled nodes. This rearrangement allows for ${\bar{\bf T}}$ to be written as four submatrices, obtained by splitting ${\bar{\bf T}}$ after the $l$th row and $l$th column as,

 \begin{equation}
{\bar{\bf T}}=
\begin{bmatrix}
 {\bf \bar{T}}_{ll} & {\bf \bar{T}}_{lu} \\
 {\bf \bar{T}}_{ul} & {\bf \bar{T}}_{uu}\\
 \end{bmatrix}.
\end{equation} 

Based on the fact that Zhu \emph{et al.}, define their label propagation algorithm as $Y \leftarrow TY$, ${\bf Y}^{U}$ can be defined as, 

\begin{equation}
\label{FP}
{\bf Y}^{U}=({\bf I}-{\bf \bar{T}}_{uu})^{-1}{\bf \bar {T}}_{ul}{\bf Y}_{L}.  
\end{equation}
More details about this label propagation approach are described in Ref. \cite{learning}. Computing ${\bf Y}^{U}$ for one pair of labeled and unlabeled node sets, $\{L\}$ and $\{U\}$, comprises the true label propagation task of one trial. To perform a null LP task, we first permute the entries of $\tilde{{\bf z}}^{L}$, and denote this permuted version as $\tilde{{\bf z}}^{L,\text{perm}}$. Just as we showed in the true label propagation task, we use $\tilde{{\bf z}}^{L,\text{perm}}$ to define a corresponding permuted version of ${\bf Y}^{U,\text{perm}}$ with $Y^{U,\text{perm}}_{ic}$ set to be 1 if node $i$ belongs to community $c$, under the permuted labels, given by $\tilde{{\bf z}}^{L,\text{perm}}$. The analogous update relationship shown in equation \ref{FP} gives ${\bf Y}^{U,\text{perm}}$. 
 
After computing ${\bf Y}^{U}$ and ${\bf Y}^{U,\text{perm}}$, the next step is to compute their corresponding entropies, $E$ and $E^{\text{perm}}$. We chose to use a cross entropy measure to account for both the attribute-based community assignment ($\tilde{\bf z}^{U}$) and the prediction under our label propagation task, ${\bf Y}^{U}$. From our attribute-based classification vector $\tilde{\bf z}$, we create the $(N-l) \times K$ indicator matrix, $\tilde{\bf Z}^{U}$,where $\tilde{\bf Z}_{ic}^{U}=1$ if node $i$ is assigned to community $c$. With this notation, we compute cross entropy $H(\tilde{\bf Z}^{U},{\bf Y}^{U})$ as,

\begin{equation}
H(\tilde{\bf Z}^{U},{\bf Y}^{U})=-\sum_{ic}\tilde{Z}_{ic}^{U}\log({Y}^{U}_{ic}).
\end{equation}
Moreover, $H(\tilde{\bf Z}^{U},{\bf Y}^{U})$ and $H(\tilde{\bf Z}^{U,\text{perm}},{\bf Y}^{U,\text{perm}})$ give $E$ and $E^{\text{perm}}$, respectively. 

We let $\mathcal{E}=\{E_{1}, E_{2}, \dots E_{T}\}$ and $\mathcal{E}_{\text{perm}}=\{E_{1}^{\text{perm}}, E_{2}^{\text{perm}}, \dots E_{T}^{\text{perm}}\}$ be the collection of entropies over the $T^{*}$ trials. 

\subsection{Computing the empirical $p$-value}
After having repeated this LP task over $T^{*}$ trials, we compute the empirical $p$-value for the test. This $p$-value is interpreted as the overlap between $\mathcal{E}$ and $\mathcal{E}_{\text{perm}}$. In the case where attributes (${\bf X}$) and connectivity (${\bf A}$) are well-aligned with connectivity, $\mathcal{E}$ and $\mathcal{E}_{\text{perm}}$ should not overlap because the entropy for the label propagation task should be very low. Alternatively, as ${\bf X}$ and ${\bf A}$ become less aligned, the entropy of the prediction from the label propagation task should be higher and hence $\mathcal{E}$ and $\mathcal{E}_{\text{perm}}$ will overlap. Then the empirical $p$-value, $p$, is calculated as,

\begin{equation}
p=P(\mathcal{E}_{\text{perm}}<\max(\mathcal{E})).
\end{equation}
 Note that this $p$-value is strictly empirical and intended to quantify the overlap between $\mathcal{E}$ and $\mathcal{E}_{\text{perm}}$.
\section{Results}
We present results on synthetic networks and on a network representation of a single cell mass cytometry CyTOF dataset. In this section, we seek to confirm that the empirical $p$-value leads to an accurate and interpretable conclusion. The results on synthetic data are useful because we have an understanding of when the $p$-value should be significant, due to our knowledge of how the data were generated. Similarly, in the single cell mass cytometry CyTOF dataset, we use particular marker features and their discriminative ability to validate our computed empirical $p$-values. 

\subsection{Synthetic Examples}
\label{sec:3A}
\begin{figure}
\centering
\begin{center}
\includegraphics[width=.5\textwidth]{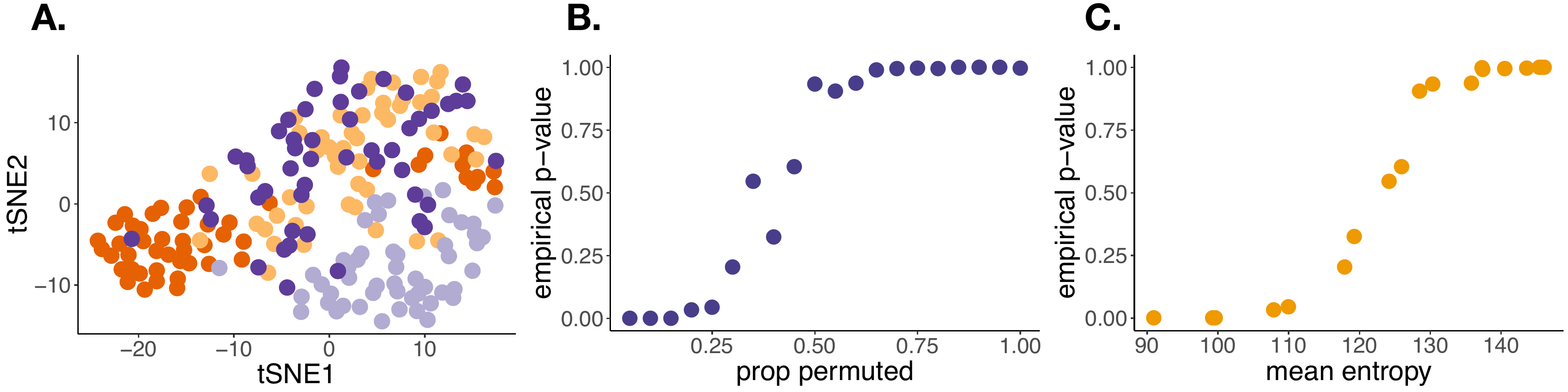}
\caption{To understand the properties of our empirical $p$-value, we generated a synthetic network, ${\bf A}$, from an SBM with $N=200$ nodes, and $K=4$ communities. The vector of continuous attributes for a node $i$, ($X_{i}$) was drawn from a multivariate Gaussian distribution parameterized by its community assignment ($z_{i}$) or $\{{\boldsymbol \mu}_{z_{i}}, {\boldsymbol \Sigma}_{z_{i}}\}$. In these experiments, we permuted varying proportions of $\tilde{{\bf z}}$ and observed the effects on entropy and empirical $p$-value. {\bf A}. We used tSNE to visualize the two dimensional projection of the 200 nodes based on attribute information. {\bf B.} We plotted the empirical $p$-value of our test as a function of the proportion of attribute-based community assignments, $\tilde{{\bf z}}$, permuted. We varied the proportion of entries of $\tilde{{\bf z}}$ permuted between 0.01 and 1 (horizontal axis). We observed decreased statistical significance (increased empirical $p$-value) with an increasing proportion of permuted labels. {\bf C.} We plotted the empirical $p$-value as a function of the mean entropy ($\mathcal{E}$) across $T^{*}=1000$ trials used to generate the entropy distributions for each experiment described in (B.). Increased entropy corresponding to a larger proportion of $\tilde{\bf z}$ permuted leads to a decreased $p$-value.}
\label{Align:Syn1}
\end{center}
\end{figure}

We sought to verify that our empirical $p$-value was capturing desirable behavior. First, we expected the $p$-value to decrease in significance as the LP entropy distribution increased in overlap with the empirical null LP entropy distribution. Second, we sought to have a $p$-value that became less significant (i.e. higher) as the correlation between attributes and network structure decreased. In Figure \ref{Align:Syn1}, we considered a network generated from a stochastic block model with $N=200$ nodes, $K=4$ communities, within-community edge probability ($p_{in}$), $p_{in}=0.6$, and between-community edge probability, ($p_{out}$), $p_{out}=0.02$. That is for a pair of nodes, $i$ and $j$, the probability of an edge existing between them is modeled as $P(A_{ij}=1)=p_{in}$ if $z_{i}=z_{j}$ and $P(A_{ij}=1)=p_{out}$ if $z_{i}\ne z_{j}$.

Associated with each node is a 3-dimensional Gaussian attribute vector, drawn from 1 of $K$ multivariate Gaussian distributions. Under this formulation, each community has its own associated multivariate Gaussian distribution. The attribute vector for a node in community $k$ is parameterized by mean ${\boldsymbol \mu}_{k}=[\mu_{1},\mu_{2},\mu_{3}]$ and covariance matrix ${\boldsymbol \Sigma}_{k}$.
 
 To generate each ${\boldsymbol \mu}_{k}=[\mu_{1},\mu_{2},\mu_{3}]$, we draw each of the $\mu_{1},\mu_{2}$ and $\mu_{3}$ from a standard Normal distribution with mean 0 and unit variance. For a community $k$, ${\boldsymbol \Sigma}_{k}$ is also the identity covariance matrix.

When performing our label propagation task, in each of the $T^{*}=1000$ trials, we used a sample of $l=100$ labeled nodes as the input to the LP task. \\
\indent First, we performed a tSNE \cite{TwoD} 2-dimensional embedding of the nodes based only on the attribute information, ${\bf X}$. This is shown in Figure \ref{Align:Syn1}A. Each point represents a node and is colored by its community assignment, ${\bf z}$. We can see that there are clearly clusters of nodes from the same community, but there is also some mixing. This implies that attributes and connectivity are not perfectly correlated. 

In this first experiment, we sought to explore how the empirical $p$-value behaved as the LP entropy distribution converged to the null LP entropy distribution. To study this, we consciously  made the label propagation task incrementally more difficult by perturbing various proportions of the initial attribute-based node-to-community assignments, $\tilde{\bf z}$. This test was implemented to verify that with a higher proportion of permuted (i.e. incorrect) entries in $\tilde{\bf z}$, the empirical entropy distributions, $\mathcal{E}$ and $\mathcal{E}_{\text{perm}}$ would have more extensive overlap. As expected, in Figure \ref{Align:Syn1}B. we observed that by permuting a larger proportion of the labels, $\tilde{\bf z}$, there was an associated increase in the empirical $p$-value (decreased significance). Here, the proportion of labels permuted in $\tilde{\bf z}$ was varied between 0.01 and 1 (horizontal axis). Next, we examined the relationship between the entropy distribution, $\mathcal{E}$ and our empirical $p$-value, $p$ in the experiments described in Figure \ref{Align:Syn1}B. In Figure \ref{Align:Syn1}C., we plot the empirical $p$-value against the mean entropy of $\mathcal{E}$ over the $T^{*}=1000$ trials. As expected, these quantities are highly related, with a higher entropy leading to a higher (less significant) empirical $p$-value.

\subsubsection{\bf Comparison to BESTest}
We used the synthetic data from the experiment described in Figure \ref{Align:Syn1} to compare our results to those obtained using BESTest \cite{peel2017ground}. Recall that BESTest is the method developed to measure the relationship between communities and a piece of node metadata in the context of a stochastic block model. This comparison is especially appropriate since the synthetic data were generated from a stochastic block model. BESTest works first by labeling the nodes according to $\tilde{\bf z}$, based on the attribute information. Under this partition of the nodes, the SBM parameters are optimized, where the maximum likelihood estimate for the connection probability between a pair of communities $r$ and $s$ is given by $\hat{\omega}_{rs}$. This maximum likelihood estimate $\hat{\omega}_{rs}$ is computed as $\hat{\omega}_{rs}=m_{rs}/n_{r}n_{s}$. Here, $m_{rs}$ is the number of edges between communities $r$ and $s$, while $n_{r}$ and $n_{s}$ are the number of nodes in communities $r$ and $s$, respectively. The entropy, $\mathcal{H}$ of this partition across the communities is computed as,

\begin{equation}
\mathcal{H}({\tilde{\bf z}})=-\frac{1}{2}[\sum_{rs}m_{rs}log \hat{\omega}_{rs}+(n_{r}n_{s}-m_{rs})\log(1-\hat{\omega}_{rs})]+O(N^{-1}). 
\end{equation}

The empirical $p$-value is computed with BESTest through a permutation test which computes $\mathcal{H}({\tilde{\bf z}}^{\text{perm}})$ many times and reports the fraction of $\mathcal{H}({\tilde{\bf z}}^{\text{perm}})<\mathcal{H}({\tilde{\bf z}})$. The BESTest entropy measure was developed in the context of a stochastic block model. While we show in subsequent experiments that our results are highly correlated with the BESTest results, our approach is developed outside of the context of the stochastic block model. We analyzed the similarity between BESTest and our label propagation approach by studying the relationship between the BESTest entropy, and our entropy and empirical $p$-value. 

\begin{figure}
\centering
\begin{center}
\includegraphics[width=.35\textwidth]{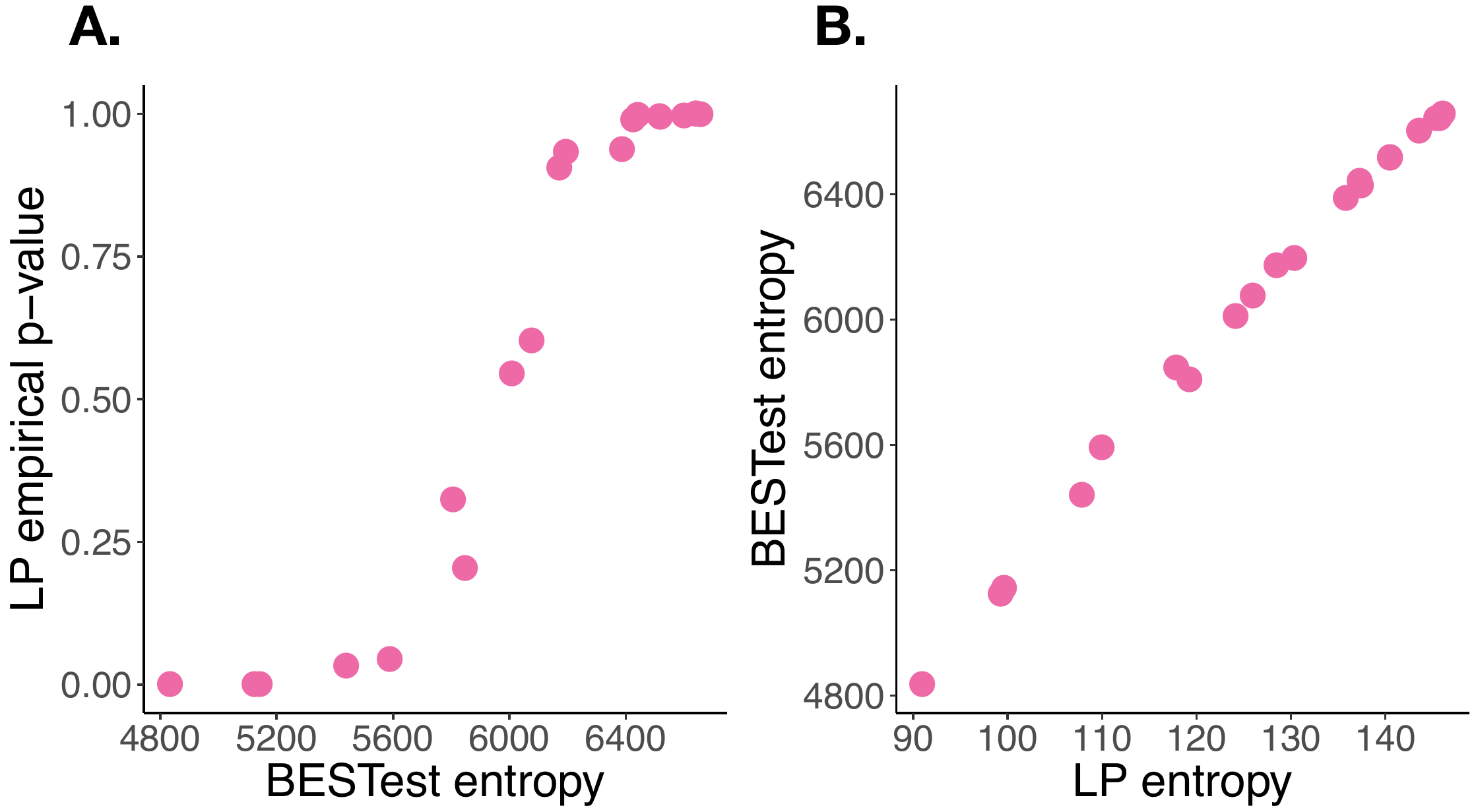}
\caption{To understand the relationship between our test and BESTest, we performed our analysis on the synthetic experiments described in section \ref{sec:3A}. In this experiment, we permute varying proportions of node-based attribute classifications, $\tilde{\bf z}$, to produce a spectrum of entropy values. From this range of entropy values that we observed with our test, we wanted to better understand the with similarity with the BESTest results. {\bf A.} There is a strong relationship between the BESTest entropy and the LP empirical $p$-value returned by our test. {\bf B.} We plotted the BESTest entropy against our label propagation entropy and observed a very strong positive correlation. This analysis demonstrates that the entropies computed by our LP test and BESTest are highly similar. }
\label{BestCompare}
\end{center}
\end{figure}

As described in section \ref{sec:3A}, our experiment involved permuting varying proportions of $\tilde{\bf z}$, which resulted in a range of entropies and hence computed empirical $p$-values. Each experiment corresponding to a particular proportion of permuted entries of $\tilde{\bf z}$ served as the attribute-based node classifications in our calculations of label propagation entropy and BESTest entropy. Note that to compute the BESTest entropies, the whole network is used. In Figure \ref{BestCompare}A. we plot our computed $p$-value against the BESTest entropy and observe that these quantities are highly related. Even more related are the BESTest and label propagation entropies, plotted against each other in Figure \ref{BestCompare}B. These analyses suggest that these tests are highly related for this particular experiment. Since we did use stochastic block models to generate our synthetic data, interesting future work could examine the relationship between these tests for more diverse types of network structures. 

\subsubsection{\bf Strength of community structure}  Given that aspects of network structure can influence label propagation results, we sought to experimentally probe how our test performed for community structure of varying strengths. We refer to a strong community structure as one that has many within-community connections and few between-community connections. In this case, communities are easy to identify, based on the prominent organizational structure. To approximate this, we considered the $p_{in}$ to $p_{out}$ ratio for a stochastic block model. As previously described, $p_{in}$ is defined as the probability of observing an edge between a pair of nodes in the same community, while $p_{out}$ is the probability of observing an edge between a pair of nodes in different communities. We expected that the entropy and empirical $p$-value would decrease with an increasing $p_{in}/p_{out}$ ratio. That is, as the community structure becomes less prominent with an increased number of connections between communities, the label propagation task should become more difficult. To study this with synthetic data, we varied the $p_{in}/p_{out}$ ratio, by considering a four community stochastic block model with values of $p_{in}$ between 0.05 and 0.45 and choosing a corresponding $p_{out}$, such that the mean degree was equal 30.  For each pair of $p_{in}$ and $p_{out}$, we generated 10 realizations from the corresponding stochastic block model. Accompanying each synthetic network was a fixed 3-dimensional attribute matrix, ${\bf X}$, where the attribute vectors for the members of community $k$ were drawn from a 3-dimensional multivariate Gaussian, parameterized by $\{{\boldsymbol \mu}_{k},{\boldsymbol \Sigma}_{k}\}$. For each synthetic network, we computed the entropy under our label propagation method and the corresponding $p$-value. 

\begin{figure}
\centering
\begin{center}
\includegraphics[width=0.5\textwidth]{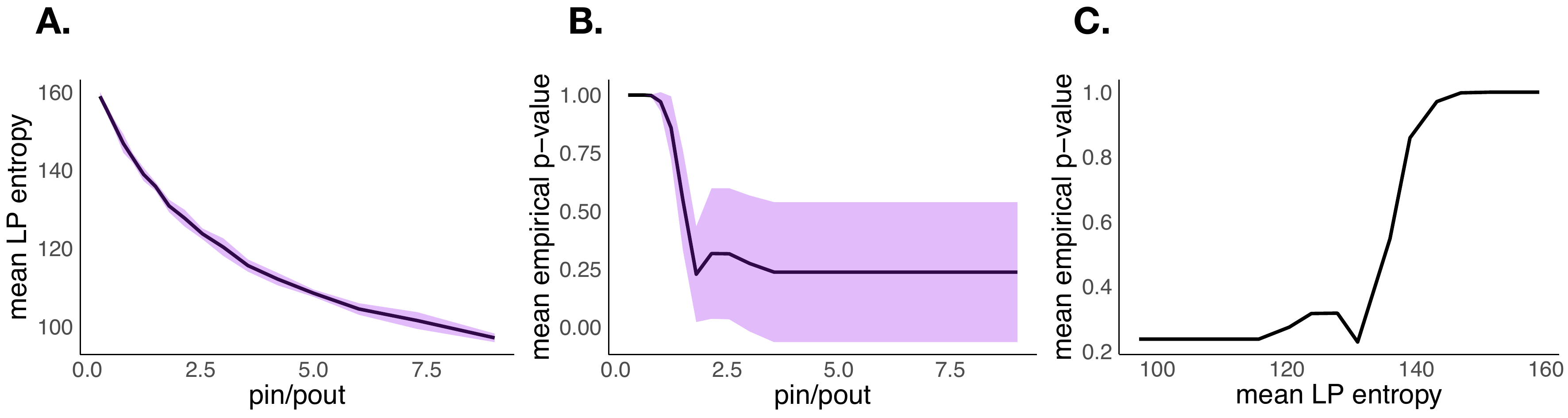}
\caption{To understand the effect of the strength of a network's community structure on our test, we generated synthetic networks from stochastic block models with various $p_{in}$ (within-community) and $p_{out}$ (between-community) parameters. Networks were generated with $p_{in}$ varying between 0.05 and 0.45 and we chose a corresponding $p_{out}$ such that the mean degree was 30. We used $p_{in}/p_{out}$ as a proxy for the strength of community, with a higher value of this ratio indicating a stronger community structure with more within-community edges and fewer between community edges. For each $p_{in}$, $p_{out}$ combination, we generated 10 synthetic network realizations. {\bf A.} We plotted the relationship between our LP entropy and $p_{in}/p_{out}$. The shaded area denotes standard deviation of the mean entropy over the 10 networks for each $p_{in}$, $p_{out}$ combination. {\bf B.} Similar to ({\bf A.}), we plotted the mean empirical $p$-value over the $T^{*}=100$ trials used to generate the entropy distributions, $\mathcal{E}$ and $\mathcal{E}_{\text{perm}}$. For large $p_{in}/p_{out}$, the empirical $p$-value became more significant. The shaded area denotes standard deviation of empirical $p$-value over the 10 networks for each $p_{in}$, $p_{out}$ combination. {\bf C.} We plotted the relationship between the mean entropy ($\mathcal{E}$) over the $T^{*}$=100 trials and 10 network realizations for each SBM parameter pair and the empirical $p$-value.}
\label{Pin}
\end{center}
\end{figure}

\indent In Figure \ref{Pin}A. we plot the mean LP entropy over the $T^{*}=100$ samples used to construct the empirical entropy distribution, $\mathcal{E}$, across the 10 network realizations for each set of $p_{in}$ and $p_{out}$. The shaded region denotes the standard deviation of the LP entropy. As the ratio between $p_{in}$ and $p_{out}$ increases, the empirical LP entropy decreases. We see a similar effect in Figure \ref{Pin}B. where we plot the empirical $p$-value as a function of the $p_{in}/p_{out}$ ratio. In this plot, the shaded region denotes the standard deviation of the empirical $p$-value. Here, a significant $p$-value (at $\alpha=0.05)$ was sometimes reached (implying attributes and connectivity are aligned) when $p_{in}/p_{out} > 3$. Finally in Figure \ref{Pin}C. we examined the relationship between the mean empirical entropy ($\mathcal{E}$) and the associated mean empirical $p$-value across the 10 network realizations generated under each parameter pair. We observe that when the LP entropy (horizontal axis) reaches 130, the mean empirical $p$-value increases (i.e. decreases in significance) very quickly.

\section{Mass Cytometry Network Example}
We applied our test to a single cell mass cytometry CyTOF dataset. Mass cytometry with CyTOF \cite{cytof} is an immunological profiling technique that gives simultaneous quantification of various immune features. The output of this technology is approximately 50 immune features profiled for each cell in a large collection of single cells. We used a freely available mass cytometry dataset, originally described in Ref. \cite{wong2015}, but pre-processed in an R tool called CytofKit \cite{cytofkit}. The dataset profiles 51 immune features across single cells on human T helper cells from peripheral blood and tonsils, which have shown to be heterogeneous within a sample. Note that consistent with the immunology literature, we also refer to these immune features as markers. To untangle the heterogeneity and infer cellular phenotypes, dimension reduction and clustering are typically applied to single cell data. In this pursuit, the objective is to cluster the single cells into predicted phenotypes, based on the measured features. A powerful way to segment the single cells into their respective phenotypes is by constructing a similarity network between the cells and clustering with community detection. This method for studying single cell data is called PhenoGraph and is described in Ref. \cite{phenoGraph}. We studied the data in an analogous way by constructing a $5$-nearest neighbor network between the cells and applying community detection to cluster them. In this representation, each node in the $k$ nearest neighbor network is a single cell and is connected to its 5 nearest neighbors, based on the pairwise Euclidean distance for the 51 measured immune features. In this example we considered a subset of 1000 single cells. After constructing the network, we predicted phenotypes by identifying communities (${\bf z}$) with the Louvain algorithm \cite{blondel}. Applying the Louvain algorithm to the network's adjacency matrix, ${\bf A}$, identified 11 communities.  As shown in Ref. \cite{cytofkit}, one further analysis after clustering the single cells is to identify features with discriminative power between inferred phenotypes. We find the application of our LP task to CyTOF data to be an appealing validation task for our algorithm because there should be a set of features (i.e. the profiled markers) that have strong discriminative ability in separating communities in the network. Note that in this context, communities in the network are also closely linked to inferred cell phenotype.  

\begin{figure}
\centering
\begin{center}
\includegraphics[width=.5\textwidth]{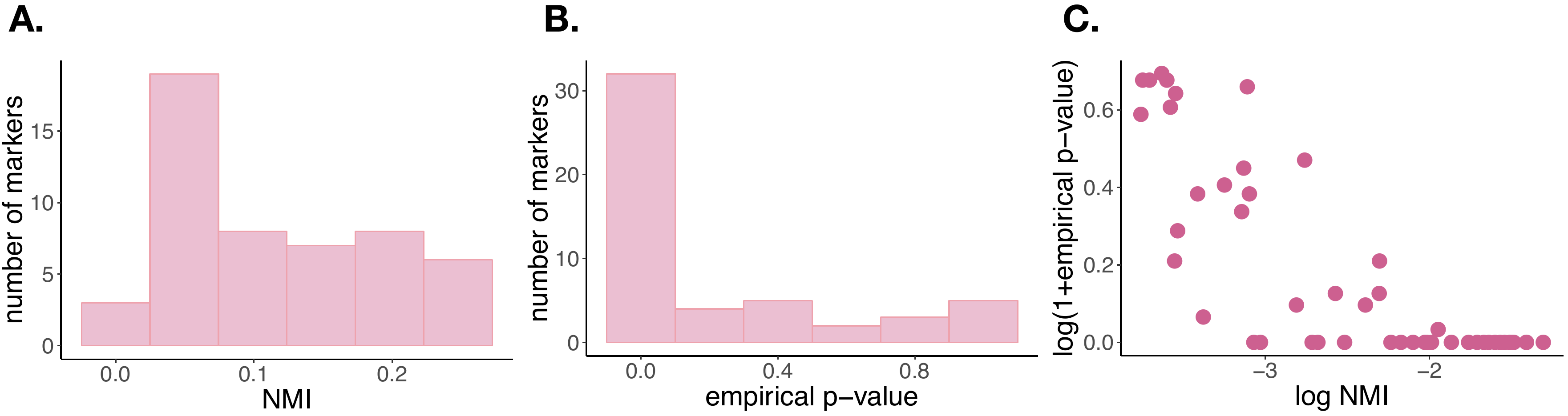}
\caption{We considered each of the 51 features in the CyTOF data and their potential to be used as discriminative markers for particular communities. We identified 11 communities (or inferred phenotypes) under the Louvain algorithm. We then created a partition, $\tilde{\bf z}$, based on each attribute in isolation. For each attribute and its induced partition of the nodes, $\tilde{\bf z}$, normalized mutual information (NMI) was used to measure the discriminative power of the marker in distinguishing network communities, or NMI($ \tilde{\bf z},{\bf z}$). We expected that our $p$-value should align with this NMI measure in that markers leading to high NMI between the induced $\tilde{\bf z}$ and ${\bf z}$ should have more significant $p$-values. {\bf A.} We used a histogram to visualize the distribution of NMI values across the 51 possible markers, with many of them leading to low NMI (between 0 and 0.1). {\bf B.} Similar to ({\bf A.}), we visualized the empirical $p$-value for the 51 possible markers. {\bf C.} We compared the relationship between the empirical $p$-value (vertical axis) and NMI($ \tilde{\bf z},{\bf z}$) (horizontal axis) across the 51 possible markers. As expected, we observed these quantities to be anti-correlated in that more significant (lower) empirical $p$-values were obtained for higher values of NMI($ \tilde{\bf z},{\bf z}$).}
\label{MarkerDist}
\end{center}
\end{figure}

The first test we performed on CyTOF $k$ nearest neighbor network was to examine how each marker feature related to the community partition, ${\bf z}$, identified with the Louvain algorithm. To understand the interplay between the $k$ nearest neighbor network structure and the measured immune features, we performed 51 separate analyses. Each analysis considered the correlation of each immune feature individually with community structure. To produce a partition of the network, $\tilde{\bf z}$ , corresponding to a single marker, $M$, we simply clustered the 1000 nodes into 1 of 11 clusters, based on the value of marker $M$. 

\begin{figure}
\begin{center}
\includegraphics[width=.5\textwidth]{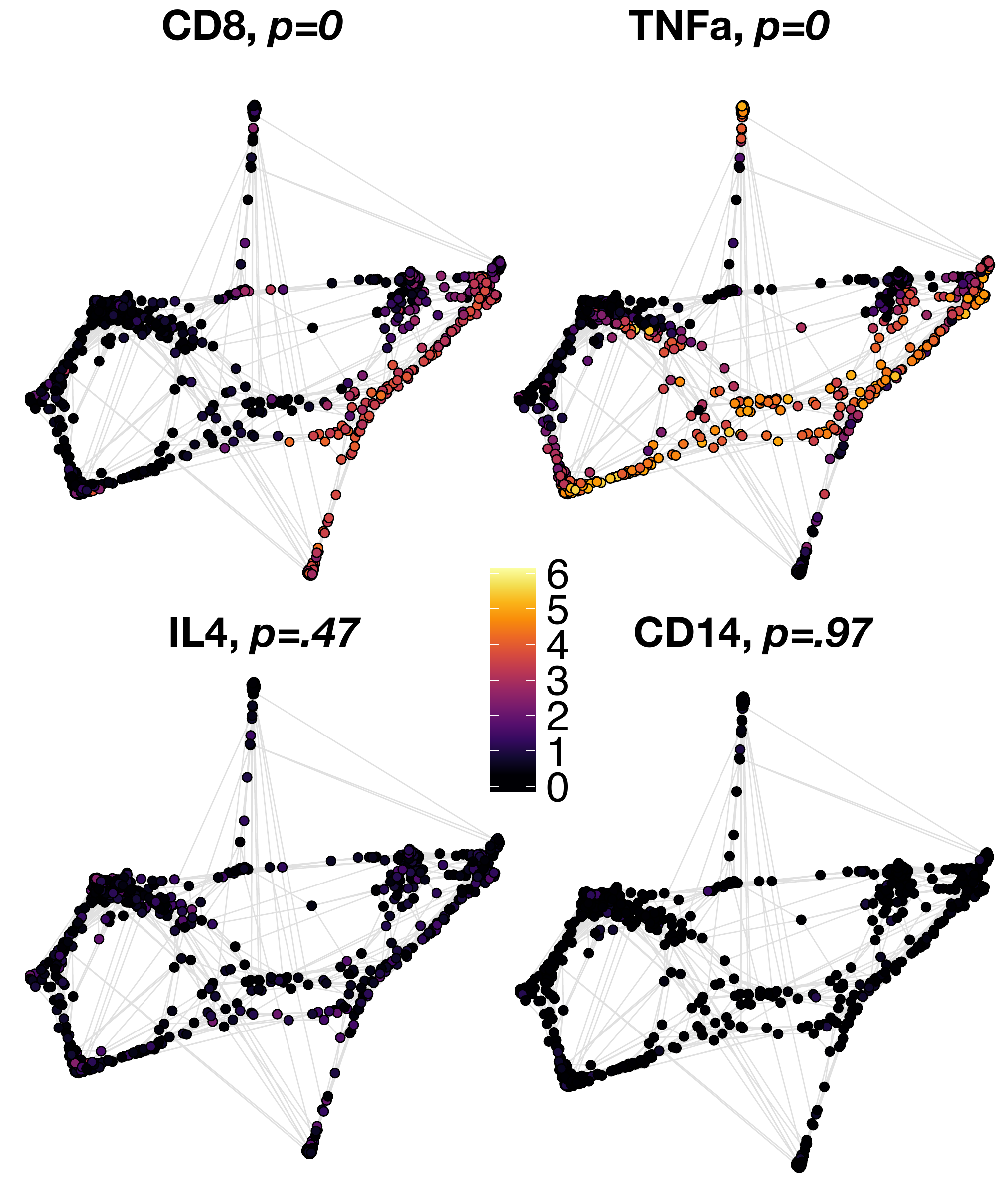}
\caption{Under our label propagation task, we chose two markers with significant $p$-values (CD8 and TNFa) and two markers with insignificant $p$-values (IL4 and CD14). We colored the nodes in the $k$-nearest neighbor network of single cells by the expression of each of these four markers. The significant markers (shown in the top row) have expression that varies across communities. Conversely, the expression of the insignificant markers does not vary across communities. This analysis suggests that our test is effective in the identification of attributes that are well aligned with community structure.}
\label{MassCy}
\end{center}
\end{figure}

Before applying our LP test to this network, we used normalized mutual information (NMI) \cite{commdeccompare} to quantify the similarity between ${\bf z}$ and $\tilde{\bf z}$. A high NMI (i.e. close to 1), indicates that the single attribute, $M$, used to create $\tilde{\bf z}$ creates a similar partition to the partition ${\bf z}$, obtained from the Louvain algorithm on just the network structure (i.e. connectivity information only). Conversely, an NMI near 0 indicates that when nodes (cells) are clustered based on the particular feature, their partition is very different to that obtained using connectivity information ({\bf A}). 

In Figure \ref{MarkerDist}A. we show the distribution of NMIs computed between ${\bf z}$ and $\tilde{\bf z}$ for each of the 51 markers. We observe a fairly broad range of marker qualities represented. Similarly, we applied our LP task for $T^{*}=30$ trials and a sample of 500 labeled nodes in each trial. Figure \ref{MarkerDist}B. shows the distribution of empirical $p$-values from our LP method. In this experiment we did not correct for multiple testing since significantly scoring features had very low empirical $p$-values (i.e. 0 or close to 0, implying no overlap between $\mathcal{E}$ and $\mathcal{E}_{\text{perm}}$). We noticed that there are approximately 30 markers with a low $p$-value (between 0 and 0.2), according to our LP test. Finally, in Figure \ref{MarkerDist}C. we examined the relationship between the NMI between ${\bf z}$ and $\tilde{\bf z}$ and the empirical $p$-value, across each of the 51 markers. As expected, these quantities are highly related, with high values of NMI corresponding to lower, more significant $p$-values. 

To visualize how particular markers correlated with communities in the network, through their induced partition, $\tilde{\bf z}$, we closely examined four different markers. Again, we used each attribute in isolation to perform an attribute-based partition of the nodes, $\tilde{\bf z}$ into 1 of 11 clusters. Two of these markers (CD8 and TNFa), had a $p$-value of 0 with our label propagation task. The other two markers (IL4 and CD14) had non-significant $p$-values of 0.47 and 0.97, respectively. In Figure \ref{MassCy}, we show the $k$ nearest neighbor networks of single cells with nodes colored by the expression for each of these four markers. Note that lighter colors indicate high expression and darker colors indicate lower expression. In the top row, we show the networks corresponding to CD8 and TNFa. We observe that expression patterns are indeed localized in the network. That is, dense clusters in the network tend to have similar marker expression. Further, this marker expression varies across communities in the network. Alternatively in the bottom row where we visualization IL4 and CD14, we observe that there is not much variability in marker expression across the network. In other words, using these markers to classify nodes would not be effective. This analysis further suggests that our empirical $p$-value can successfully identify markers with discriminative power because they are well-correlated with particular communities.

\begin{figure}
\begin{center}
\includegraphics[width=.3\textwidth]{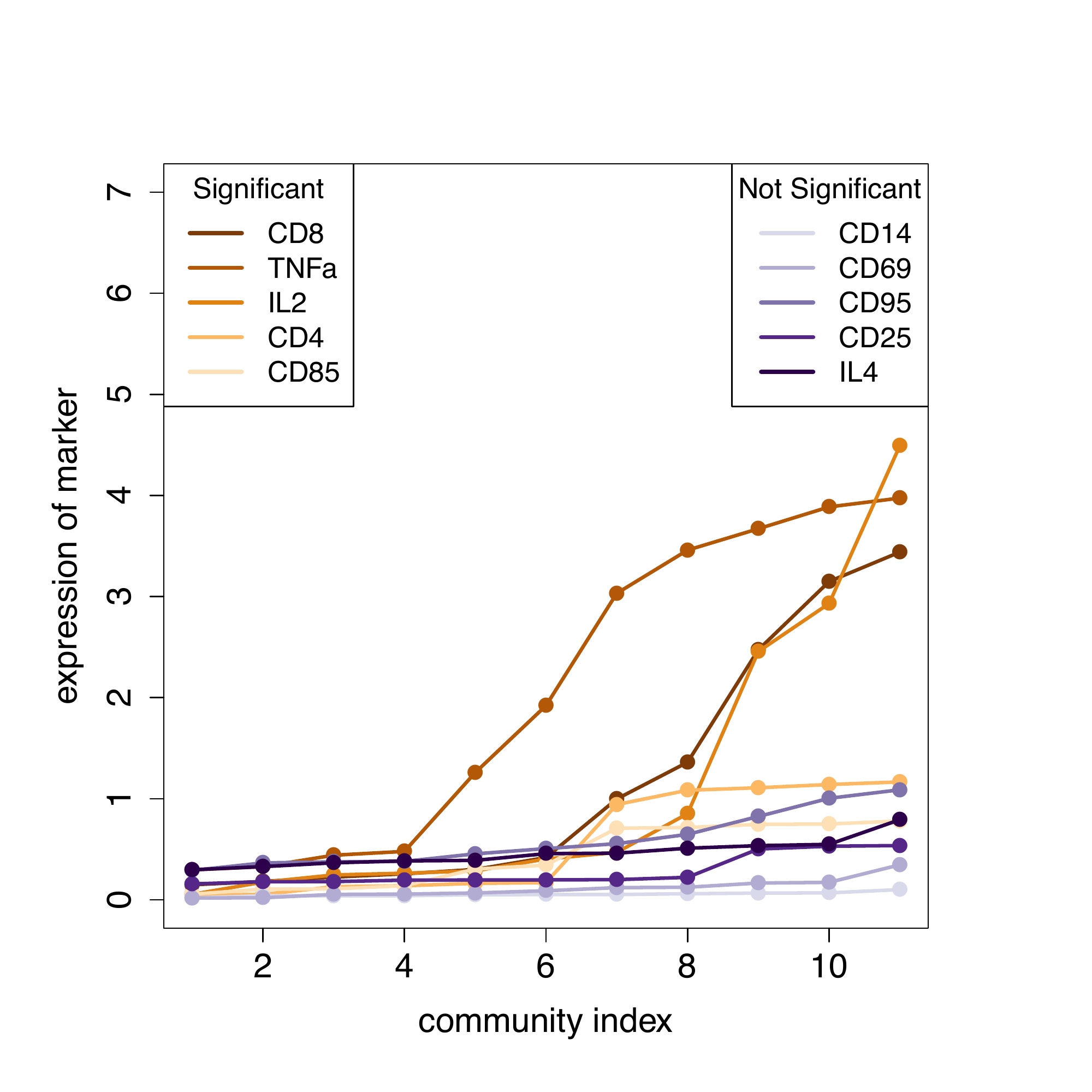}
\caption{ We computed the empirical $p$-values induced by the partition $\tilde{\bf z}$ for each of the 51 markers in isolation. Using our test, we looked closely at 5 of the most and least significant markers under the computed the empirical $p$-value. Since a quality marker in this case is said to be one that induces a labeling of the nodes, $\tilde{\bf z}$, similar to the result obtained under ${\bf z}$, we expect the expression of such a marker to vary across communities. In this plot, we show the expression of each marker as a function of the community index. The family of orange-colored lines correspond to the top 5 significant markers (according to empirical $p$-value). From all of these lines, the expression varies across communities. Conversely, looking at the markers with low significance under our test, expression is relatively constant across all communities.}
\label{MarkerRank}
\end{center}
\end{figure}

As an final experiment, we sought to see if the markers with significant empirical $p$-values (implying that they are effective in distinguishing cellular phenotypes) did indeed vary across communities in the network, through their induced partition, $\tilde{\bf z}$. To do this, we selected 10 markers from the 51 measured features of the CyTOF data. In particular, we looked at the 5 of the most and least significant markers, in terms of the computed empirical $p$-value. For each of these 10 markers, we computed the mean marker expression across each of the 11 communities identified by applying the Louvain algorithm applied to the network's adjacency matrix, ${\bf A}$. in Figure \ref{MarkerRank}, we then plotted the mean marker expression across communities for the 5 most and least significant markers. The least significant markers are shown in the family of blue lines and are relatively static across each of the 10 communities. In contrast, the orange family of lines corresponds to the markers for the more significant features and do vary across communities. Since a marker with a significant low empirical $p$-value should correlate well with communities, this is the pattern we expected. The 5 poorly ranked markers clearly do not correlate with communities because their expression is constant across all communities.

\section{Conclusion}
In this paper, we introduced a label propagation based approach to determine how closely attributes align with network connectivity. Over $T^{*}$ trials, the label propagation task uses a subset of nodes labeled according to attribute information to predict the labels for a set of unlabeled nodes. The label propagation task we adopt returns a probability distribution for each of the unlabeled nodes over the possible communities. The empirical $p$-value of our test is computed by comparing the empirical entropy distributions across the $T^{*}$ trials from our label propagation task, and a permuted label propagation task, denoted by $\mathcal{E}$ and $\mathcal{E}_{\text{perm}}$, respectively. The intuition is that if attributes are well aligned with network connectivity patterns, then the label propagation task should produce results that are more certain, and hence have lower entropy. Our results indicate that the computed entropy and empirical $p$-value are behaving as expected on synthetic examples, where we designed the experiments in a way that we knew how well the attributes and connectivity correlated. We also show that our test is useful in the identification of important marker features for distinguishing communities in the single cell mass cytometry $k$ nearest neighbor network. Here, features (markers) with low empirical $p$-value are features that vary across communities and hence give insight into immune features that distinguish communities or cell subsets. 

As future work, one might examine how the entropy and empirical $p$-value relate to different communities identified using different algorithms or approaches. Future work could also benefit from the analysis of the number of trials ($T^{*}$) and the optimal number of nodes in each of the labeled and unlabeled node sets that lead to the most statistically stable and meaningful results. Finally, similar to how we detected particular marker features that were aligned with the identified communities, perhaps we can use our tool as a feature selection method that can can be used to create meaningful network representations of data. 

\section*{Acknowledgments}
This work was supported by the National Science Foundation under award \#1610762.

\bibliographystyle{ACM-Reference-Format}
\bibliography{ABib}


\begin{thebibliography}{23}


\ifx \showCODEN    \undefined \def \showCODEN     #1{\unskip}     \fi
\ifx \showDOI      \undefined \def \showDOI       #1{#1}\fi
\ifx \showISBNx    \undefined \def \showISBNx     #1{\unskip}     \fi
\ifx \showISBNxiii \undefined \def \showISBNxiii  #1{\unskip}     \fi
\ifx \showISSN     \undefined \def \showISSN      #1{\unskip}     \fi
\ifx \showLCCN     \undefined \def \showLCCN      #1{\unskip}     \fi
\ifx \shownote     \undefined \def \shownote      #1{#1}          \fi
\ifx \showarticletitle \undefined \def \showarticletitle #1{#1}   \fi
\ifx \showURL      \undefined \def \showURL       {\relax}        \fi
\providecommand\bibfield[2]{#2}
\providecommand\bibinfo[2]{#2}
\providecommand\natexlab[1]{#1}
\providecommand\showeprint[2][]{arXiv:#2}

\bibitem[\protect\citeauthoryear{Bendall, Nolan, Roederer, and
  Chattopadhyay}{Bendall et~al\mbox{.}}{2012}]%
        {cytof}
\bibfield{author}{\bibinfo{person}{Sean~C Bendall}, \bibinfo{person}{Garry~P
  Nolan}, \bibinfo{person}{Mario Roederer}, {and} \bibinfo{person}{Pratip~K
  Chattopadhyay}.} \bibinfo{year}{2012}\natexlab{}.
\newblock \showarticletitle{A deep profiler's guide to cytometry}.
\newblock \bibinfo{journal}{\emph{Trends in immunology}} \bibinfo{volume}{33},
  \bibinfo{number}{7} (\bibinfo{year}{2012}), \bibinfo{pages}{323--332}.
\newblock


\bibitem[\protect\citeauthoryear{Benson, Gleich, and Leskovec}{Benson
  et~al\mbox{.}}{2016}]%
        {benson}
\bibfield{author}{\bibinfo{person}{Austin~R Benson}, \bibinfo{person}{David~F
  Gleich}, {and} \bibinfo{person}{Jure Leskovec}.}
  \bibinfo{year}{2016}\natexlab{}.
\newblock \showarticletitle{Higher-order organization of complex networks}.
\newblock \bibinfo{journal}{\emph{Science}} \bibinfo{volume}{353},
  \bibinfo{number}{6295} (\bibinfo{year}{2016}), \bibinfo{pages}{163--166}.
\newblock


\bibitem[\protect\citeauthoryear{Blondel, Guillaume, Lambiotte, and
  Lefebvre}{Blondel et~al\mbox{.}}{2008}]%
        {blondel}
\bibfield{author}{\bibinfo{person}{Vincent~D Blondel},
  \bibinfo{person}{Jean-Loup Guillaume}, \bibinfo{person}{Renaud Lambiotte},
  {and} \bibinfo{person}{Etienne Lefebvre}.} \bibinfo{year}{2008}\natexlab{}.
\newblock \showarticletitle{Fast unfolding of communities in large networks}.
\newblock \bibinfo{journal}{\emph{Journal of statistical mechanics: theory and
  experiment}} \bibinfo{volume}{2008}, \bibinfo{number}{10}
  (\bibinfo{year}{2008}), \bibinfo{pages}{P10008}.
\newblock


\bibitem[\protect\citeauthoryear{Chen, Lau, Wong, Newell, Poidinger, and
  Chen}{Chen et~al\mbox{.}}{2016}]%
        {cytofkit}
\bibfield{author}{\bibinfo{person}{Hao Chen}, \bibinfo{person}{Mai~Chan Lau},
  \bibinfo{person}{Michael~Thomas Wong}, \bibinfo{person}{Evan~W Newell},
  \bibinfo{person}{Michael Poidinger}, {and} \bibinfo{person}{Jinmiao Chen}.}
  \bibinfo{year}{2016}\natexlab{}.
\newblock \showarticletitle{Cytofkit: a bioconductor package for an integrated
  mass cytometry data analysis pipeline}.
\newblock \bibinfo{journal}{\emph{PLoS computational biology}}
  \bibinfo{volume}{12}, \bibinfo{number}{9} (\bibinfo{year}{2016}),
  \bibinfo{pages}{e1005112}.
\newblock


\bibitem[\protect\citeauthoryear{Combe, Largeron, G{\'e}ry, and
  Egyed-Zsigmond}{Combe et~al\mbox{.}}{2015}]%
        {ilouvain}
\bibfield{author}{\bibinfo{person}{David Combe}, \bibinfo{person}{Christine
  Largeron}, \bibinfo{person}{Mathias G{\'e}ry}, {and}
  \bibinfo{person}{El{\H{o}}d Egyed-Zsigmond}.}
  \bibinfo{year}{2015}\natexlab{}.
\newblock \showarticletitle{I-Louvain: An Attributed Graph Clustering Method}.
\newblock In \bibinfo{booktitle}{\emph{Advances in Intelligent Data Analysis
  XIV}}. \bibinfo{publisher}{Springer}, \bibinfo{pages}{181--192}.
\newblock


\bibitem[\protect\citeauthoryear{Danon, Diaz-Guilera, Duch, and Arenas}{Danon
  et~al\mbox{.}}{2005}]%
        {commdeccompare}
\bibfield{author}{\bibinfo{person}{Leon Danon}, \bibinfo{person}{Albert
  Diaz-Guilera}, \bibinfo{person}{Jordi Duch}, {and} \bibinfo{person}{Alex
  Arenas}.} \bibinfo{year}{2005}\natexlab{}.
\newblock \showarticletitle{Comparing community structure identification}.
\newblock \bibinfo{journal}{\emph{Journal of Statistical Mechanics: Theory and
  Experiment}} \bibinfo{volume}{2005}, \bibinfo{number}{09}
  (\bibinfo{year}{2005}), \bibinfo{pages}{P09008}.
\newblock


\bibitem[\protect\citeauthoryear{Fortunato and Hric}{Fortunato and
  Hric}{2016}]%
        {fortu2}
\bibfield{author}{\bibinfo{person}{Santo Fortunato} {and}
  \bibinfo{person}{Darko Hric}.} \bibinfo{year}{2016}\natexlab{}.
\newblock \showarticletitle{Community detection in networks: A user guide}.
\newblock \bibinfo{journal}{\emph{Physics Reports}}  \bibinfo{volume}{659}
  (\bibinfo{year}{2016}), \bibinfo{pages}{1--44}.
\newblock


\bibitem[\protect\citeauthoryear{Hric, Peixoto, and Fortunato}{Hric
  et~al\mbox{.}}{2016}]%
        {hric}
\bibfield{author}{\bibinfo{person}{Darko Hric}, \bibinfo{person}{Tiago~P
  Peixoto}, {and} \bibinfo{person}{Santo Fortunato}.}
  \bibinfo{year}{2016}\natexlab{}.
\newblock \showarticletitle{Network structure, metadata, and the prediction of
  missing nodes and annotations}.
\newblock \bibinfo{journal}{\emph{Physical Review X}} \bibinfo{volume}{6},
  \bibinfo{number}{3} (\bibinfo{year}{2016}), \bibinfo{pages}{031038}.
\newblock


\bibitem[\protect\citeauthoryear{Levine, Simonds, Bendall, Davis, El-ad,
  Tadmor, Litvin, Fienberg, Jager, Zunder, et~al\mbox{.}}{Levine
  et~al\mbox{.}}{2015}]%
        {phenoGraph}
\bibfield{author}{\bibinfo{person}{Jacob~H Levine}, \bibinfo{person}{Erin~F
  Simonds}, \bibinfo{person}{Sean~C Bendall}, \bibinfo{person}{Kara~L Davis},
  \bibinfo{person}{D~Amir El-ad}, \bibinfo{person}{Michelle~D Tadmor},
  \bibinfo{person}{Oren Litvin}, \bibinfo{person}{Harris~G Fienberg},
  \bibinfo{person}{Astraea Jager}, \bibinfo{person}{Eli~R Zunder},
  {et~al\mbox{.}}} \bibinfo{year}{2015}\natexlab{}.
\newblock \showarticletitle{Data-driven phenotypic dissection of AML reveals
  progenitor-like cells that correlate with prognosis}.
\newblock \bibinfo{journal}{\emph{Cell}} \bibinfo{volume}{162},
  \bibinfo{number}{1} (\bibinfo{year}{2015}), \bibinfo{pages}{184--197}.
\newblock


\bibitem[\protect\citeauthoryear{Maaten and Hinton}{Maaten and Hinton}{2008}]%
        {TwoD}
\bibfield{author}{\bibinfo{person}{Laurens van~der Maaten} {and}
  \bibinfo{person}{Geoffrey Hinton}.} \bibinfo{year}{2008}\natexlab{}.
\newblock \showarticletitle{Visualizing data using t-SNE}.
\newblock \bibinfo{journal}{\emph{Journal of machine learning research}}
  \bibinfo{volume}{9}, \bibinfo{number}{Nov} (\bibinfo{year}{2008}),
  \bibinfo{pages}{2579--2605}.
\newblock


\bibitem[\protect\citeauthoryear{Mucha, Richardson, Macon, Porter, and
  Onnela}{Mucha et~al\mbox{.}}{2010}]%
        {muchacommunity}
\bibfield{author}{\bibinfo{person}{Peter~J Mucha}, \bibinfo{person}{Thomas
  Richardson}, \bibinfo{person}{Kevin Macon}, \bibinfo{person}{Mason~A Porter},
  {and} \bibinfo{person}{Jukka-Pekka Onnela}.} \bibinfo{year}{2010}\natexlab{}.
\newblock \showarticletitle{Community structure in time-dependent, multiscale,
  and multiplex networks}.
\newblock \bibinfo{journal}{\emph{science}} \bibinfo{volume}{328},
  \bibinfo{number}{5980} (\bibinfo{year}{2010}), \bibinfo{pages}{876--878}.
\newblock


\bibitem[\protect\citeauthoryear{Newman}{Newman}{2006}]%
        {newmanmodularity}
\bibfield{author}{\bibinfo{person}{Mark~EJ Newman}.}
  \bibinfo{year}{2006}\natexlab{}.
\newblock \showarticletitle{Modularity and community structure in networks}.
\newblock \bibinfo{journal}{\emph{Proceedings of the National Academy of
  Sciences}} \bibinfo{volume}{103}, \bibinfo{number}{23}
  (\bibinfo{year}{2006}), \bibinfo{pages}{8577--8582}.
\newblock


\bibitem[\protect\citeauthoryear{Newman and Clauset}{Newman and
  Clauset}{2016}]%
        {clauset}
\bibfield{author}{\bibinfo{person}{Mark~EJ Newman} {and} \bibinfo{person}{Aaron
  Clauset}.} \bibinfo{year}{2016}\natexlab{}.
\newblock \showarticletitle{Structure and inference in annotated networks}.
\newblock \bibinfo{journal}{\emph{Nature Communications}}  \bibinfo{volume}{7}
  (\bibinfo{year}{2016}), \bibinfo{pages}{11863}.
\newblock


\bibitem[\protect\citeauthoryear{Peel, Larremore, and Clauset}{Peel
  et~al\mbox{.}}{2017}]%
        {peel2017ground}
\bibfield{author}{\bibinfo{person}{Leto Peel}, \bibinfo{person}{Daniel~B
  Larremore}, {and} \bibinfo{person}{Aaron Clauset}.}
  \bibinfo{year}{2017}\natexlab{}.
\newblock \showarticletitle{The ground truth about metadata and community
  detection in networks}.
\newblock \bibinfo{journal}{\emph{Science Advances}} \bibinfo{volume}{3},
  \bibinfo{number}{5} (\bibinfo{year}{2017}), \bibinfo{pages}{e1602548}.
\newblock


\bibitem[\protect\citeauthoryear{Perozzi and Akoglu}{Perozzi and
  Akoglu}{2018}]%
        {perozziAttribute}
\bibfield{author}{\bibinfo{person}{Bryan Perozzi} {and} \bibinfo{person}{Leman
  Akoglu}.} \bibinfo{year}{2018}\natexlab{}.
\newblock \showarticletitle{Discovering Communities and Anomalies in Attributed
  Graphs: Interactive Visual Exploration and Summarization}.
\newblock \bibinfo{journal}{\emph{ACM Trans. Knowl. Discov. Data}}
  \bibinfo{volume}{12}, \bibinfo{number}{2}, Article \bibinfo{articleno}{24}
  (\bibinfo{date}{Jan.} \bibinfo{year}{2018}), \bibinfo{numpages}{40}~pages.
\newblock
\showISSN{1556-4681}
\urldef\tempurl%
\url{https://doi.org/10.1145/3139241}
\showDOI{\tempurl}


\bibitem[\protect\citeauthoryear{Shai, Stanley, Granell, Taylor, and
  Mucha}{Shai et~al\mbox{.}}{2017}]%
        {shaicase}
\bibfield{author}{\bibinfo{person}{Saray Shai}, \bibinfo{person}{Natalie
  Stanley}, \bibinfo{person}{Clara Granell}, \bibinfo{person}{Dane Taylor},
  {and} \bibinfo{person}{Peter~J Mucha}.} \bibinfo{year}{2017}\natexlab{}.
\newblock \showarticletitle{Case studies in network community detection}.
\newblock \bibinfo{journal}{\emph{arXiv preprint arXiv:1705.02305}}
  (\bibinfo{year}{2017}).
\newblock


\bibitem[\protect\citeauthoryear{Snijders and Nowicki}{Snijders and
  Nowicki}{1997}]%
        {sbmOrig}
\bibfield{author}{\bibinfo{person}{Tom~AB Snijders} {and}
  \bibinfo{person}{Krzysztof Nowicki}.} \bibinfo{year}{1997}\natexlab{}.
\newblock \showarticletitle{Estimation and prediction for stochastic
  blockmodels for graphs with latent block structure}.
\newblock \bibinfo{journal}{\emph{Journal of classification}}
  \bibinfo{volume}{14}, \bibinfo{number}{1} (\bibinfo{year}{1997}),
  \bibinfo{pages}{75--100}.
\newblock


\bibitem[\protect\citeauthoryear{Stanley, Bonacci, Kwitt, Niethammer, and
  Mucha}{Stanley et~al\mbox{.}}{2018}]%
        {stanley2018}
\bibfield{author}{\bibinfo{person}{Natalie Stanley}, \bibinfo{person}{Thomas
  Bonacci}, \bibinfo{person}{Roland Kwitt}, \bibinfo{person}{Marc Niethammer},
  {and} \bibinfo{person}{Peter~J Mucha}.} \bibinfo{year}{2018}\natexlab{}.
\newblock \showarticletitle{Stochastic Block Models with Multiple Continuous
  Attributes}.
\newblock \bibinfo{journal}{\emph{arXiv preprint arXiv:1803.02726}}
  (\bibinfo{year}{2018}).
\newblock


\bibitem[\protect\citeauthoryear{Wong, Chen, Narayanan, Lin, Anicete, Kiaang,
  De~Lafaille, Poidinger, and Newell}{Wong et~al\mbox{.}}{2015}]%
        {wong2015}
\bibfield{author}{\bibinfo{person}{Michael~T Wong}, \bibinfo{person}{Jinmiao
  Chen}, \bibinfo{person}{Sriram Narayanan}, \bibinfo{person}{Wenyu Lin},
  \bibinfo{person}{Rosslyn Anicete}, \bibinfo{person}{Henry Tan~Kun Kiaang},
  \bibinfo{person}{Maria Alicia~Curotto De~Lafaille}, \bibinfo{person}{Michael
  Poidinger}, {and} \bibinfo{person}{Evan~W Newell}.}
  \bibinfo{year}{2015}\natexlab{}.
\newblock \showarticletitle{Mapping the diversity of follicular helper T cells
  in human blood and tonsils using high-dimensional mass cytometry analysis}.
\newblock \bibinfo{journal}{\emph{Cell reports}} \bibinfo{volume}{11},
  \bibinfo{number}{11} (\bibinfo{year}{2015}), \bibinfo{pages}{1822--1833}.
\newblock


\bibitem[\protect\citeauthoryear{Xie and Szymanski}{Xie and Szymanski}{2011}]%
        {LabProp}
\bibfield{author}{\bibinfo{person}{Jierui Xie} {and}
  \bibinfo{person}{Boleslaw~K Szymanski}.} \bibinfo{year}{2011}\natexlab{}.
\newblock \showarticletitle{Community detection using a neighborhood strength
  driven label propagation algorithm}. In \bibinfo{booktitle}{\emph{Network
  Science Workshop (NSW), 2011 IEEE}}. IEEE, \bibinfo{pages}{188--195}.
\newblock


\bibitem[\protect\citeauthoryear{Yang and Leskovec}{Yang and Leskovec}{2013}]%
        {bigclam}
\bibfield{author}{\bibinfo{person}{Jaewon Yang} {and} \bibinfo{person}{Jure
  Leskovec}.} \bibinfo{year}{2013}\natexlab{}.
\newblock \showarticletitle{Overlapping community detection at scale: a
  nonnegative matrix factorization approach}. In
  \bibinfo{booktitle}{\emph{Proceedings of the sixth ACM international
  conference on Web search and data mining}}. ACM, \bibinfo{pages}{587--596}.
\newblock


\bibitem[\protect\citeauthoryear{Yang, McAuley, and Leskovec}{Yang
  et~al\mbox{.}}{2013}]%
        {cesna}
\bibfield{author}{\bibinfo{person}{Jaewon Yang}, \bibinfo{person}{Julian
  McAuley}, {and} \bibinfo{person}{Jure Leskovec}.}
  \bibinfo{year}{2013}\natexlab{}.
\newblock \showarticletitle{Community detection in networks with node
  attributes}. In \bibinfo{booktitle}{\emph{Data mining (ICDM), 2013 ieee 13th
  international conference on}}. IEEE, \bibinfo{pages}{1151--1156}.
\newblock


\bibitem[\protect\citeauthoryear{Zhu and Ghahramani}{Zhu and
  Ghahramani}{2002}]%
        {learning}
\bibfield{author}{\bibinfo{person}{Xiaojin Zhu} {and} \bibinfo{person}{Zoubin
  Ghahramani}.} \bibinfo{year}{2002}\natexlab{}.
\newblock \showarticletitle{Learning from labeled and unlabeled data with label
  propagation}.
\newblock  (\bibinfo{year}{2002}).
\newblock


\end{thebibliography}

\end{document}